\documentclass{epl}
\usepackage{amssymb}

\title{Phase transition in the modified fiber bundle model}
\shorttitle{Phase transitions in the modified fiber bundle model}

\author{Beom Jun Kim\thanks{E-mail: \email{beomjun@ajou.ac.kr}}}
\shortauthor{B.J. Kim}
\institute { Department of Molecular Science
        and Technology, Ajou University, Suwon 442-749, Korea} 


\begin{document}

\maketitle

\begin{abstract}

We extend the standard fiber bundle model (FBM) with the local
load sharing in such a way that the conservation of the
total load is relaxed when an isolated fiber is broken.
In this modified FBM in one dimension (1D), it is
revealed that the model exhibits a well-defined phase
transition at a finite nonzero value of the load, which
is in contrast to the standard 1D FBM.
The modified FBM defined in the  Watts-Strogatz network is also 
investigated, and found is the
existences of two distinct transitions: one discontinuous 
and the other continuous.  The effects of the long-range 
shortcuts are also discussed.

\end{abstract}

The fiber bundle model (FBM) was introduced to explain a variety of phenomena
related with the problems of damage spread~\cite{classic}. More
specifically, consider a system composed of a bundle  of mingled fibers exposed
to a common stress along the direction of fibers. It is natural to assume that
each fiber has a certain value of the capacity, or the threshold value of the
load, and if the load  on the fiber exceeds the capacity allowed to it, the
fiber is broken.  Since the broken fiber stops functioning, its neighbor fibers
should carry the load which was assigned to the broken one~\cite{lls}.  The damage caused
by the breaking of a single fiber may propagate over its local neighbors and
then stop spreading.  Another interesting possibility is the existence of an
avalanche or a cascading failure, i.e., the original local damage spreads over
the whole system, causing the complete breakdown of the total fibers. Recent
blackout in the United States can also be understood as the cascading failures
of power cables and power stations caused by the spread of the overload
breakdowns of elements in power grid.

Recently, the subject of complex networks is the one of the most 
studied research area, not only
for its intrinsic interest but also for its promising
broad future applicability~\cite{ref:network}. When a dynamic system
is defined on a complex network, the emergent behavior may
totally differ from the one on the local regular array which has
been the standard arena of statistical physics. For example,
the phase transitions of the Ising and the $XY$ models on  complex 
networks have been shown to belong to the mean-field universality 
class~\cite{isingXY}.
The spread of damage on networks, which is not described by
the equilibrium Hamiltonian, also draws much interest
in relation to the epidemic spread~\cite{epidemic} as well as 
the overload breakdowns of networks subject to local~\cite{morenoepl}
or nonlocal~\cite{holme} loads.

The fiber bundle model in this work  on a general network structure is defined
 as follows:
\begin{enumerate}
\item To each vertex in the network of the size $N$ the capacity 
$c_i$ for the $i$th vertex is  assigned  at random  following a given distribution function (in most cases we use
the uniform distribution with $c_i \in [0,1]$). 
\item The system is then exposed to a total load $N\sigma$, and thus each
vertex should carry the same load $\sigma_i = \sigma$
(we in this work $\sigma = 0$ is used as an initial condition).
\item Examine the breaking condition $\sigma_i > c_i$ for each $i$
and if the inequality is satisfied the vertex $i$ is broken and its load $\sigma_i$ is equally
transfered to its surviving neighbor vertices. For example, if the 
vertex $i$ has $k_i$ surviving neighbors the vertex $j$, which is
a neighbor of $i$, has now new load: 
$\sigma_j \rightarrow \sigma_j + \sigma_i/k_i$. Repeat this step until
there is no more breaking.
\item
Increase the total load $N\sigma \rightarrow N(\sigma + d\sigma)$.
The total load should be carried by 
unbroken surviving vertices, the number of which is $N_s$, and accordingly
each vertex is assigned the new load $\sigma_i \rightarrow \sigma_i +
(N/N_s)d\sigma$. Note that we denote the local load on the vertex $i$
as $\sigma_i$ while $\sigma$ is the total load divided by $N$.
\end{enumerate} 
The above procedures are identical to the standard FBM with the local
load sharing (LLS) in the literature~\cite{lls}. However,   
our modified FBM in the present work differs as follows:
\begin{itemize}
\item When an isolated vertex, which has no surviving neighbor vertices,
is overload broken, we assume that
the load which has been carried by this vertex is not transferred 
to other vertices. On the other hand, in the standard FBM with
LLS the load transfers via edges, not via chain of surviving vertices,
and consequently the load of the broken isolated fiber can still
be transferred to other (suitably defined) closely located vertices.

\end{itemize}
Our modified FBM appears to be reasonable 
in itself since the standard FBM assumes that the load can be
transferred to vertices which are not connected to the broken vertex
through a chain of {\it surviving} vertices.
In contrast, our model allows only the load transfer across
the actually existing surviving vertices.
One disadvantage of our modified FBM
is that the total load applied to the system 
is not constant in time.  It might be difficult to find 
the actual physical realization of our model in the original context of the
FBM. However, in abstract networks such as Internet
one may not rule out the possibility that when an isolated 
vertex is broken its load does not spread to other vertices but just disappears.
It should be noted that even in our modified FBM, the total load is resumed 
when each vertex is assigned the new load as 
$\sigma_i \rightarrow \sigma_i + (N/N_s)d\sigma$  (see step 4) since the
breaking of an isolated vertex also decreases $N_s$. Accordingly, the
total load conservation condition is relaxed only during the spread of
damage in step 3.

We repeat the above described procedure for our modified version of
FBM and measure
the ratio $\rho \equiv N_s/N$ as a function of the 
load $\sigma$ averaged over $10^4 \sim 10^5$ different capacity
distributions and different network structures. 
In most numerical simulations, we use $d\sigma = 0.001$, which
has been numerically confirmed to be small enough, and 
network sizes are $N=128, 256, \cdots, 524 \; 288$ (or from $N=2^7$ to
$2^{19}$).

\begin{figure}
\twofigures[height=4.8cm]{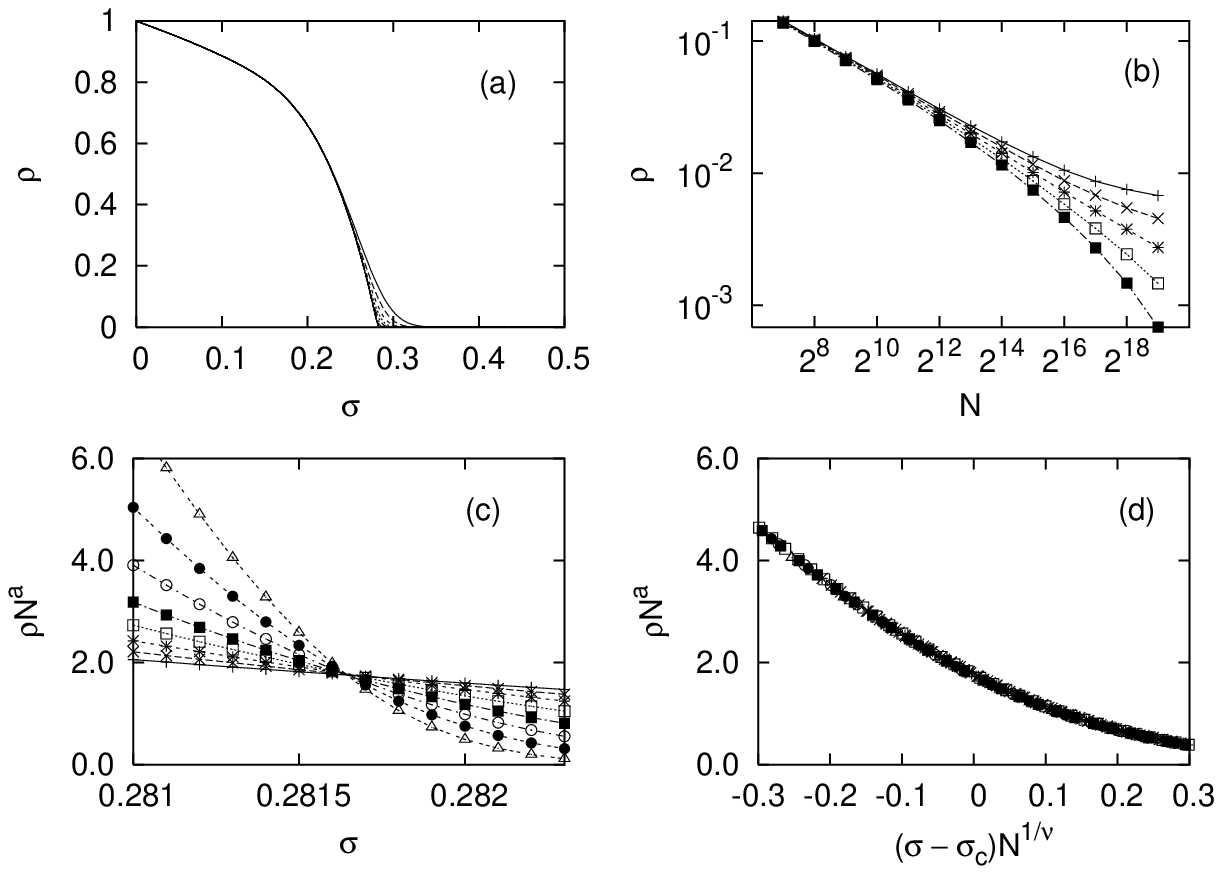}{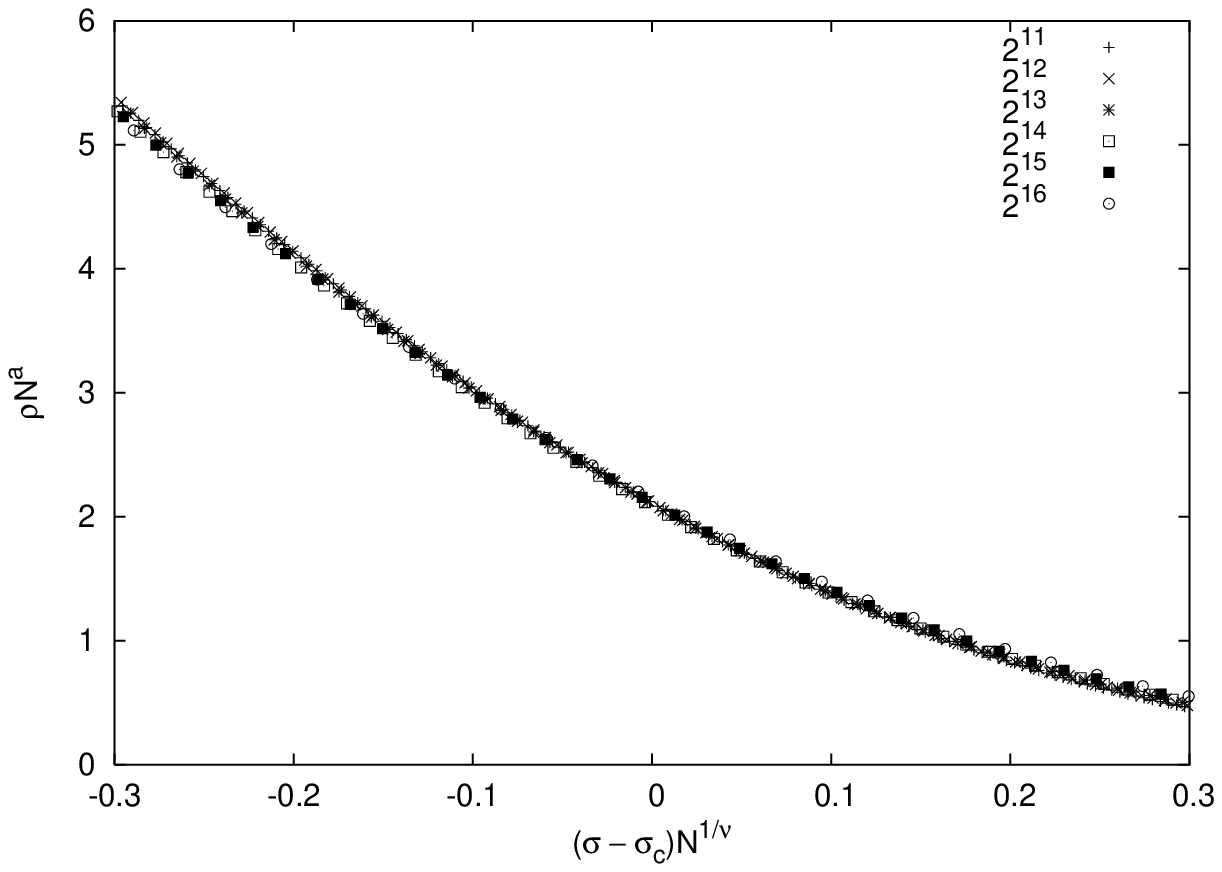}
\caption{The modified fiber bundle model on the 1D regular lattice 
with couplings up to the next nearest neighbors with the 
uniform distribution of the capacity $c_i \in [0,1]$.
(a) Density $\rho$ of unbroken fibers versus the load $\sigma$ 
for $N=2^7, 2^8, \cdots, 2^{19}$ (from top to bottom).
As $\sigma$ is increased, the network begins to break up, resulting in 
$\rho = 0$ at sufficiently large values of $\sigma$.
(b) $\rho$ versus $N$ for $\sigma = 0.2812, 0.2814, 0.2816, 0.2818$, 
and 0.2820 (from top to bottom). Around $\sigma \approx 0.2817$, $\rho$
decays algebraically with $N$, manifesting the existence of the 
continuous phase transition. (c) Precise determination of $\sigma_c$
from the finite-size scaling form~(\ref{eq:scaling}). Determined are
$a=0.50(1)$ and $\sigma_c = 0.28165(2)$.  
(Curves correspond to $N = 2^{12}, 2^{13}, \cdots, 2^{19}$, respectively, 
from bottom to top on the left-hand side of the crossing point).
(d) Scaling collapse of all data points in (c) through the use
of the finite-size scaling~(\ref{eq:scaling}); $\nu = 2.0(1)$ is obtained.
}
\label{fig:p0}
\caption{The FBM on the 1D regular lattice with connections
up to the next nearest neighbors with the Gaussian distribution
of the capacity (the mean and the variance are fixed the same as in Fig.~\ref{fig:p0}). 
Finite-size scaling (compare with Fig.~\ref{fig:p0} for the 
uniform distribution) yields $\sigma_c = 0.27363(2)$ and critical
exponents $a=\beta/\nu = 0.50(1)$ and $\nu = 2.0(1)$, identical
to the uniform distribution.
}
\label{fig:p0gauss}
\end{figure}

We first consider the local regular one-dimensional (1D) network with
the connection range $r$. 
For various values of $r$, including $r=1$
which corresponds to the usual 1D lattice, we observe the
qualitatively the same behavior.
In Fig.~\ref{fig:p0}(a) for $r=2$, it is shown that 
at a certain value of the load $\sigma$ the ratio $\rho$
between the number of working fibers and the total number of
fibers vanishes beyond some critical value $\sigma_c$. As the
system size $N$ becomes larger 
[$N=2^7, 2^8, \cdots, 2^{19} (=524 \; 288)$ in Fig.~\ref{fig:p0}(a)], the phase transition becomes clearer.
It is interesting to note that the existence of the phase transition
at nonzero $\sigma_c$ is completely different from the
well-established result of $\sigma_c = 0$ for the standard FBM 
in the thermodynamic limit~\cite{lls}. Simple change of the
conservation law of the total load yields the dramatic change
of the universality class.

We in Fig.~\ref{fig:p0} (b) show $\rho$ as a function of $N$
in log scales at various values of 
$\sigma = 0.2812, 0.2814, 0.2816, 0.2818$, and 0.2820 (from top to bottom). 
For $\sigma > \sigma_c \approx 0.2817$, $\rho (N)$  shows a downward
curvature, implying $\rho \rightarrow 0$ in the thermodynamic limit, while
$\rho(N)$ bends upward for $\sigma < \sigma_c$, suggesting that $\rho(N)$
saturates to a finite value as $N$ becomes larger.
Separating these behaviors, there exists a well-defined $\sigma_c$ at which
$\rho$ as a function of $N$ decays algebraically, and the system does
not possess any length scale except $N$ at $\sigma = \sigma_c$.  This
is commonly found behavior in the system with a continuous phase transition.

For a more precise determination of $\sigma_c$, we use the standard
finite-size scaling method with the scaling assumption:
\begin{equation}  \label{eq:scaling}
\rho (\sigma, N) = N^{-a} f( \bigl (\sigma - \sigma_c) N^{1/\nu} \bigr ) ,
\end{equation} 
where $f(x)$ is the scaling function and the exponent $\nu$
describes the divergence of the correlation length $\xi$ [or more generally,
the correlation volume for the case when the length is ill-defined (see
discussions in, e.g., Ref.~\cite{isingXY})]
near the critical point, i.e., $\xi \sim |\sigma - \sigma_c|^{-\nu}$.
From the definition of the critical exponent $\beta$,
$\rho \sim (\sigma_c - \sigma)^\beta$ for $\sigma < \sigma_c$
in the thermodynamic limit,
we obtain the relation $a = \beta/\nu$.
Figure~\ref{fig:p0}(c) is for a precise determination of $\sigma_c$
from the finite-size scaling~(\ref{eq:scaling}): $\rho N^a$ is shown
to have a unique crossing point at $\sigma = \sigma_c = 0.28165(2)$
with the exponent $a=\beta/\nu = 0.50(1)$.
Once $\sigma_c$ and the exponent $a$ are determined, one can again
use the finite-size scaling~(\ref{eq:scaling}) to make the all data
points in Fig.~\ref{fig:p0}(c) collapse to a single smooth curve as displayed
in Fig.~\ref{fig:p0}(d) with the determination of $\nu = 2.0(1)$.

The above results of the existence of the continuous phase transition
with critical exponents $\nu \approx 2$ and $\beta \approx 1$ is
expected to be universal regardless of the detailed form of the
distribution function of the capacity. We in Fig.~\ref{fig:p0gauss} show
the finite-size scaling plot for the same FBM but with 
the Gaussian distribution of the capacity instead. For comparisons,
the same values of the  mean  and the variance as ones with 
the uniform distribution in Fig.~\ref{fig:p0} are used.
Confirmed is the existence of the continuous phase transition
with the identical critical exponents. 
Although we have
not used the commonly used Weibull distribution, we believe
that the universality class should be identical for the
various capacity distribution functions. 

\begin{figure}
\onefigure{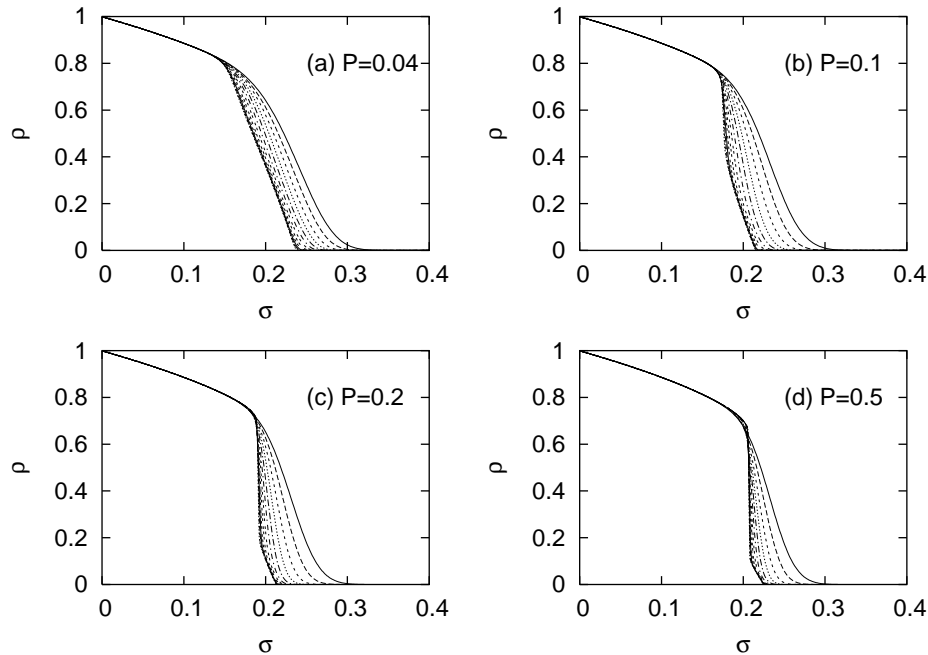}
\caption{The FBM on the Watts-Strogatz small-world networks
with the connection range $r=2$ and at various rewiring probabilities
$P=$ (a) 0.04, (b) 0.1, (c) 0.2, and (d) 0.5. Except (a), all 
show interesting transition behaviors: the density of working
fibers $\rho$ drops down abruptly at a certain value of the
load $\sigma_1$ and then approaches $\rho = 0$ continuously
at $\sigma_2$. Networks with the sizes $N = 2^7, 2^8, \cdots 2^{19}$
(from top to bottom in each plot) 
and the uniform capacity distribution are used.
}
\label{fig:pall}
\end{figure}

\begin{figure}
\twofigures[height=4.8cm]{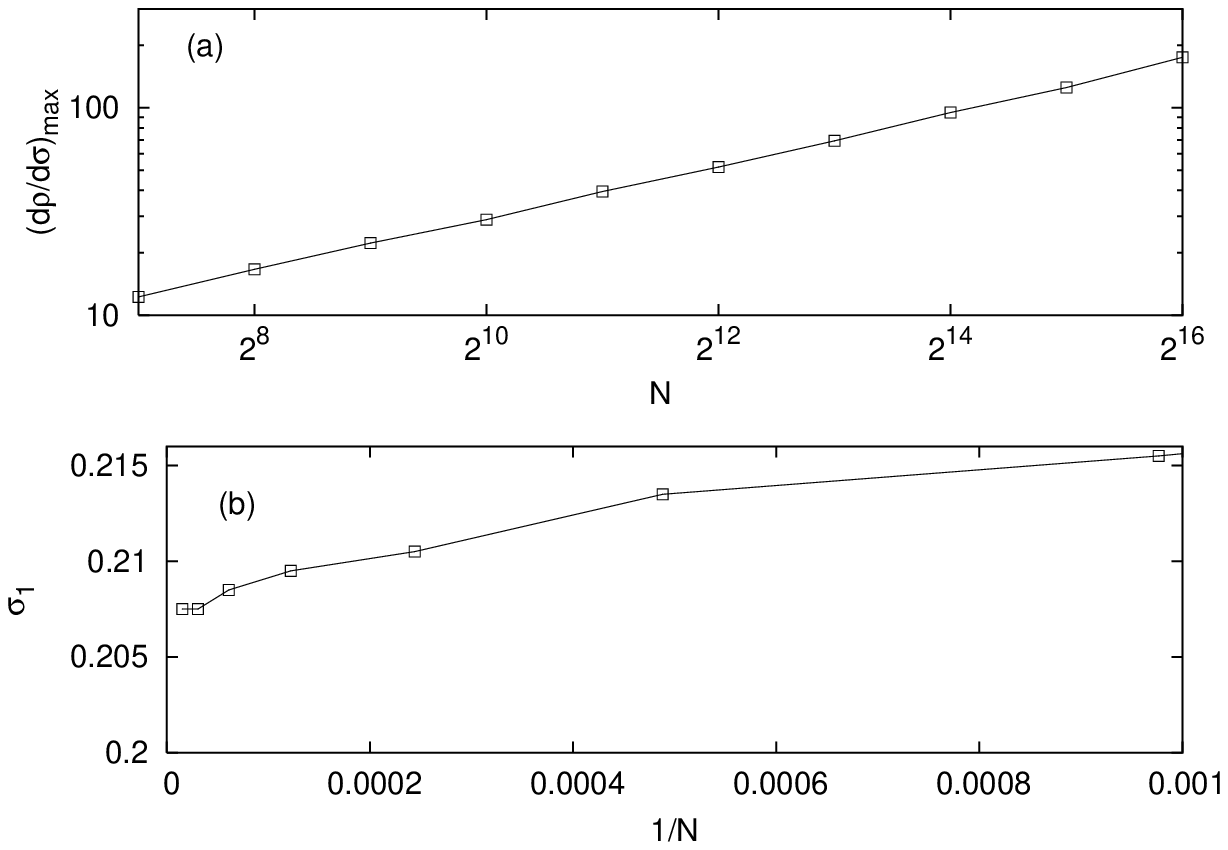}{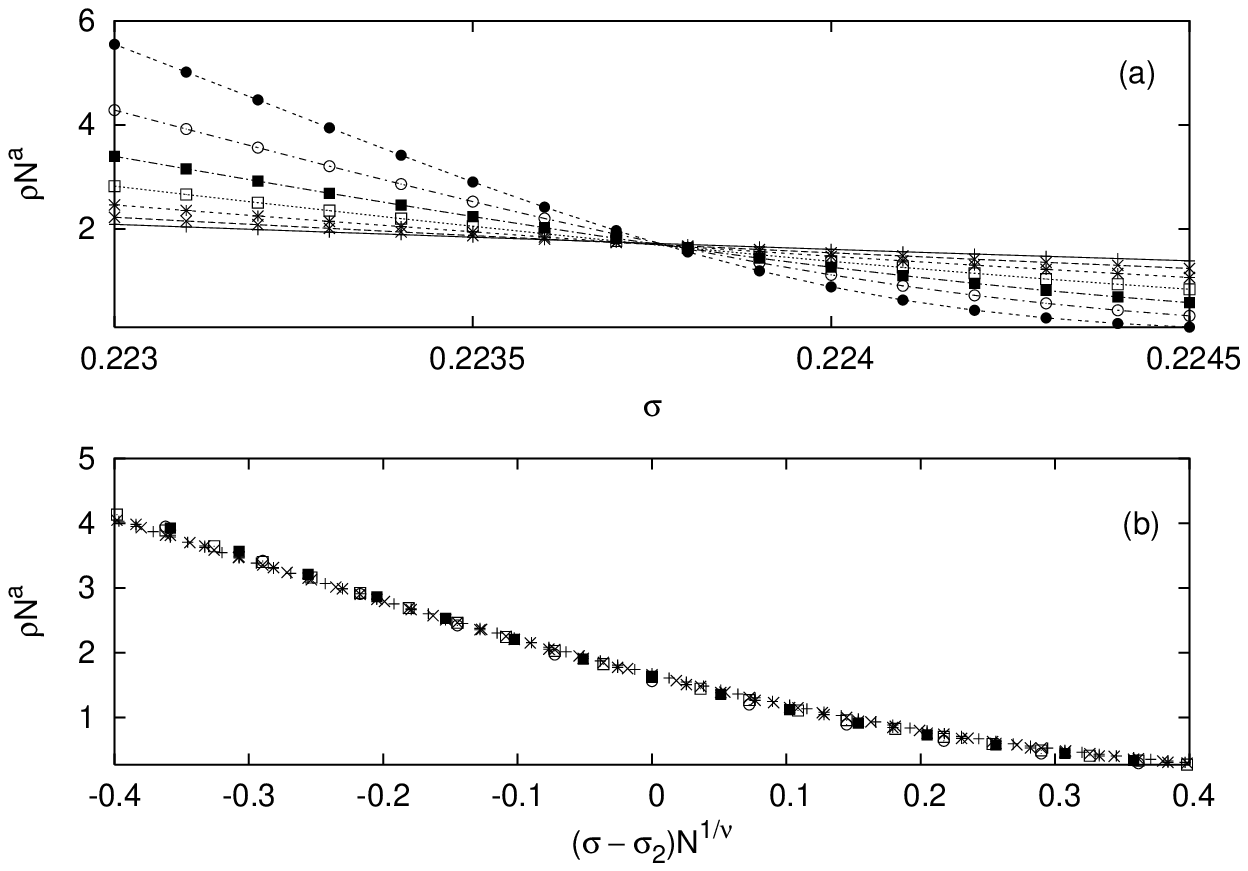}
\caption{The FBM on the WS network with the rewiring probability $P = 0.5$. (a)
The maximum of the derivative of $\rho(\sigma)$ with respect to $\sigma$ versus
the system size $N$. The divergence of the derivative implies the discontinuous
transition nature in the thermodynamic limit. $\sigma_1$ is defined as the
value of $\sigma$ when this maximum occurs. (b) The location $\sigma_1$ of the
derivative maximum versus $1/N$. In the thermodynamic limit $\sigma_1 \approx
0.207$ is estimated.
}
\label{fig:p0.5disc}
\caption{
The continuous phase transition in the
FBM on the WS network with $P=0.5$. (a) $\rho N^a$ versus $\sigma$
at various sizes $N=2^{13}, \cdots 2^{19}$ with $a=0.5$ crosses
at a unique crossing point, resulting in the estimation $\sigma_2 
= 0.22376(2)$ as the critical point of the second phase transition
of the continuous nature. (b) Finite-size scaling collapse~(\ref{eq:scaling})
of data points in (a). The critical exponents $a = \beta/\nu = 0.5,
\nu = 2.0$ (and thus $\beta = 1.0$) result in the scaling
collapse.
}
\label{fig:p0.5cont}
\end{figure}

We next study the FBM on the Watts-Strogatz (WS) networks, which is generated
following the standard procedure in Ref.~\cite{WS}.  The two important
parameters in the WS network are the connection range $r$, ($r=2$ is used here)
and the rewiring probability $P$ controlling the number of long-range
shortcuts. As soon as $P$ has a nonzero value, it is well-known that the
network undergoes so-called small-world transition that the network diameter
increases logarithmically with the network size $N$.  In Fig.~\ref{fig:pall}
$\rho$ versus $\sigma$ is shown at $P=$ (a) 0.04, (b) 0.1, (c) 0.2, and (d)
0.5. At sufficiently large values of $P$, it is clearly demonstrated that the
system has two phase transitions at $\sigma_1$ and $\sigma_2$ ($\sigma_1 <
\sigma_2$): At $\sigma_1$, $\rho$ appears to drop abruptly to a nonzero finite
value and as $\sigma$ is further increased, $\rho$ eventually vanishes beyond
$\sigma_2$.

To investigate the nature of the first transition at $\sigma_1$, 
we compute $d\rho / d\sigma$ as a function of $\sigma$ and then measure
at which value of $\sigma$ the derivative takes the maximum value.
In Fig.~\ref{fig:p0.5disc}(a), the maximum value of the derivative
$(d\rho / d\sigma)_{\rm max}$ is plotted as a function of the system 
size $N$: The power-law divergence 
$(d\rho / d\sigma)_{\rm max} \sim N^{0.4}$ clearly suggests that the
transition nature is indeed discontinuous in the thermodynamic limit
of $N \rightarrow \infty$. The location of the
derivative maximum, $\sigma_1$, is displayed as a function of $1/N$
in Fig.~\ref{fig:p0.5disc}(b); $\sigma_1 \approx 0.207$ is concluded
in the thermodynamic limit.
We then apply the finite-size scaling analysis to the regime
where $\rho$ approaches zero continuously (see Fig.~\ref{fig:p0.5cont}).
The unique crossing point in the $\rho N^a$ versus $\sigma$ at
various $N$ in Fig.~\ref{fig:p0.5cont}(a) as well as the smooth
scaling collapse in Fig.~\ref{fig:p0.5cont}(b) is again observed
with the identical values of critical exponents, i.e, $a \approx 0.5$, 
$\nu \approx 2.0$, and $\beta \approx 1.0$.
It is noteworthy that the FBM on the scale-free network of Barab{\'a}si and
Albert (BA)~\cite{BA} has  shown to have a different transition
nature~\cite{morenoepl}, i.e., a discontinuous phase transition 
without the subsequent continuous transition 
as in the WS network considered here. Consequently,
as soon as the network undergoes the global spread of the damage, no 
fiber survives the disaster in the BA network, while in the WS network 
fibers which survived the discontinuous transition are subject to the second
continuous transition at  a higher load. 

In the context of complex
networks, the size of giant component has been more frequently measured
since it can detect the global breakdown of the information flow between
two arbitrarily chosen vertices. In the viewpoint of the universality,
the use of the different order parameters is expected to
change neither the nature of the phase transition nor the values of critical
exponents, which are the main interest in this work. The density of 
surviving fibers in the WS network can also have a practical meaning 
if one considers the real fiber bundles in which most inter-fiber
couplings are local while some of them are long-ranged.

\begin{figure}
\onefigure{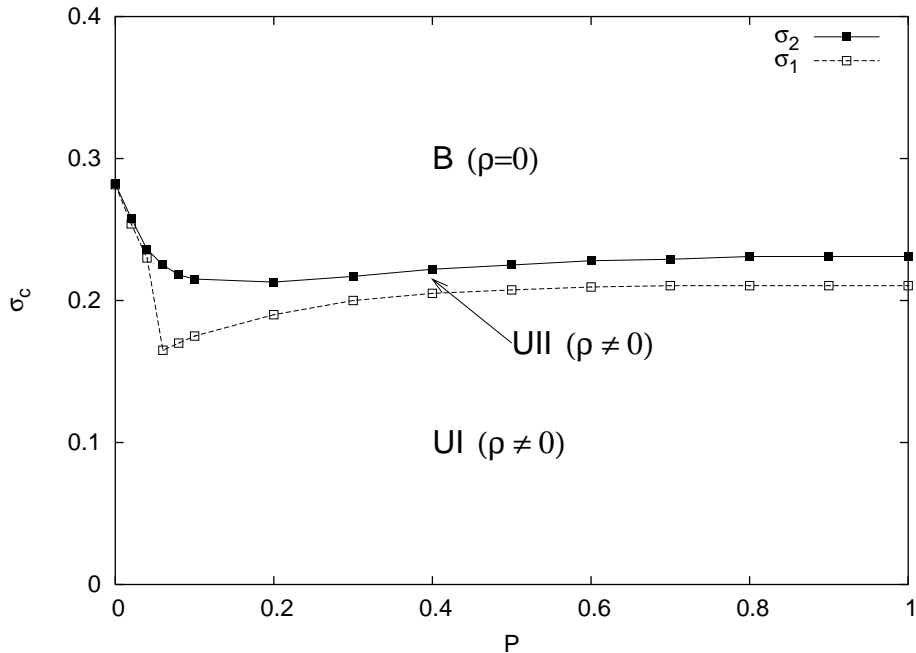}
\caption{Phase diagram of the FBM on the WS network with the
connection range $r=2$. The points (filled squares) on the 
full line are obtained
from the finite-size scaling analysis for the continuous phase
transition (see Fig.~\ref{fig:p0.5cont}) and the points (empty squares)
on the dotted line are from the maximum
position of $d\rho/d\sigma$ extrapolated to $N \rightarrow \infty$
(see Fig.~\ref{fig:p0.5disc}). There exist three phases: two unbroken
phases with $\rho \neq 0$ (UI and  UII), and completely broken phase
with $\rho =0$ (B). In UI, most fibers are still working with higher
value of $\rho \gtrsim 0.7$, while in UII only small fraction (but
still nonzero) of fibers hold together to endure the external load.
The transition between UI and UII is discontinuous
while between UII and B the nature of the transition is continuous.
For sufficiently small values of $P (\lesssim 0.04)$, it appears that there exists
only one transition from UI to B.
}
\label{fig:phd}
\end{figure}

Figure~\ref{fig:phd} summarizes the phase diagram obtained in the
present work. There exist in general two phase transition
lines separating three different phases (two unbroken phases UI and
UII with $\rho \neq 0$ and the broken phase B with $\rho = 0$): 
$\sigma_1$ is estimated from the maximum of $d\rho/d\sigma$ 
while $\sigma_2$ is obtained from the standard finite-size
scaling analysis applied for the continuous phase transition.
As the load $\sigma$ is increased from zero, the system only suffers from
locally isolated damage spreading (the phase UI in Fig.~\ref{fig:phd})
until the first discontinuous transition
point $\sigma_1$ is reached. 
As $\sigma$ crosses $\sigma_1$ from below, the system
undergoes abrupt change, probably in the form of the cascading failures
or an avalanche, and the number of fibers still alive decays discontinuously. 
Interestingly,
the existence of the long-range shortcuts not only makes the discontinuous
transition happen,  but also help the system to avoid
complete disaster, i.e, the global spread of damage does not destroy
all fibers in the system but there still remain a finite fraction of fibers which
survived the first transition at $\sigma_1$ (the phase UII in Fig.~\ref{fig:phd}).  
Further increase of the load then eventually destroys
all fibers resulting in the $\rho = 0$ state (phase B in Fig.~\ref{fig:phd}).
At sufficiently small values of $P$, it appears that
the system has only one phase transition which is continuous in
nature: At $P \lesssim 0.04$, it is found that the maximum 
of $d\rho/d\sigma$ and the continuous phase transition occur simultaneously.
At $P \gtrsim 0.06$, the system starts to have well-separated two
phase transitions, and the increase of $P$ appears to make the system more
strong (both $\sigma_1$ and $\sigma_2$ increase with $P$ in the
intermediate region).

In summary, we in this work have extended the existing model of fiber bundles
in such a way that the total load conservation condition is relaxed. For a pure
1D regular network, our modified FBM has been found to show well-defined
continuous phase transition at a finite nonzero value of the load $\sigma_c$
with the critical exponents $\beta \approx 1.0$ and $\nu \approx 2.0$. This is
in a sharp contrast to the standard FBM where $\sigma_c = 0$ has been
confirmed in the thermodynamic limit.  It has also been verified that the
universality class of the phase transition does not change when different
distribution function for the capacity is tried.  For the Watts-Strogatz
network, as the rewiring probability $P$ is increased, the system undergoes two
phase transitions at $\sigma_1$ and $\sigma_2$. At $\sigma_1$ the density
$\rho$ of working fibers decreases discontinuously to a nonzero finite value.
At $\sigma_2 (> \sigma_1)$, $\rho$ approaches continuously zero, with the phase
transition with the identical universality class ($\beta \approx 1.0$ and $\nu
\approx 2.0$).  The roles played by the existence of long-range shortcuts have
been shown to be two-fold: On the one hand, long-range shortcuts make the fiber
breaking transition abrupt, helping the damage spread across the system more
easily. On the other hand, the long-range shortcuts prohibit the system from
being completely destroyed, and  even after the global propagation of damage
(at $\sigma_1$), there still remain a finite fraction of fibers still working. 

The author thanks D.H. Kim for useful discussions.
This work has been supported by the Korea Science
and Engineering Foundation through Grant 
No. R14-2002-062-01000-0 and Hwang-Pil-Sang research
fund in Ajou University. 
Numerical works have been
performed on the computer cluster Iceberg at Ajou University.

\end{document}